\documentstyle[aps,epsf,eqsecnum,feynmp,graphics,amsfonts, amssymb, latexsym]{revtex}
\begin{document}
\title{The Highest-Derivative Version of
Variational Perturbation Theory}
\author{B.~Hamprecht\index{HAMPRECHT, B.} and 
A.~Pelster\index{PELSTER, A.}}
\address{Institut f\"ur Theoretische Physik, Freie Universit\"at Berlin,\\
Arnimallee 14, D-14195 Berlin, Germany\\
E-mails: bodo.hamprecht@physik.fu-berlin.de, 
axel.pelster@physik.fu-berlin.de}

\maketitle

\begin{abstract}
We systematically investigate different versions of
variational perturbation theory by forcing not only the first or
second but also higher derivatives of the approximant with respect
to the variational parameter to vanish. The choice of the highest derivative
version turns out to be the most successful one for approximating
the ground-state energy of the 
anharmonic oscillator\index{anharmonic oscillator}. It is therefore used
to determine the critical exponent\index{critical exponent} 
$\alpha$ of the
specific heat\index{specific heat} 
in superfluid ${}^4$He in agreement with the value measured
in recent space shuttle experiments.
\end{abstract}

\section{Introduction}
The perturbative treatment of quantum-statistical or field-theoretical
problems renders in general results in the form of divergent infinite
power series in some coupling constant $g$. Typically, the
coefficients of these series grow factorially with
alternating signs leading to a zero convergence radius. Various
resummation schemes may be applied to obtain finite
results for all values of the coupling constant $g$, even in the
strong-coupling limit\index{strong-coupling limit} 
$g \rightarrow \infty$ (for an overview see
Chap.~16 of the book of Kleinert and 
Schulte-Frohlinde\index{SCHULTE-FROHLINDE, V.}~\cite{HAMPVerena} and the 
references therein). Most successful is a recent systematic development
by Kleinert~\cite{HAMPKleinert1}, extending the
variational method of Feynman\index{FEYNMAN, R.P.} and
Kleinert\index{KLEINERT, H.}~\cite{HAMPFeynman} which was set up for
calculating the effective classical potential in quantum statistics.
It has been thoroughly
tested for the ground-state energy of the 
anharmonic oscillator\index{anharmonic oscillator} and
shown to converge exponentially fast and uniform to the correct
result~\cite{HAMPJanke1}.  This was encouraging enough to
apply the method also to divergent series which arise
from renormalizing the $\phi^4$-theory of critical
phenomena\index{critical phenomena}~\cite{HAMPVerena,HAMPKleinert2,HAMPKleinert3,HAMPKleinert4,HAMPKleinert5,HAMPKleinert6},
where the perturbation coefficients are available
up to six and partly to seven loops 
in $d=3$~\cite{HAMPNickel1,HAMPSokolov,HAMPNickel2} and up
to five loops in $d=4-\epsilon$ dimensions~\cite{HAMPFive}.  The
method yielded finite results with a smooth dependence on the order.
Furthermore, the theoretical results are in excellent 
agreement with the only
experimental value available so far with an appropriate accuracy, the
critical exponent\index{critical exponent} 
$\alpha$ governing the behaviour of the specific
heat\index{specific heat} 
near the superfluid phase transition of
${}^4$He~\cite{HAMPKleinert7,HAMPLipa}.

Let us briefly recall the method.
Consider some function $f(g)$ which is perturbatively obtained
for a small coupling constant $g$ as the divergent weak-coupling
series
\begin{eqnarray}
\label{HAMPEX1}
f ( g ) = \sum_{n=0}^{\infty} c_n g^n \, ,
\end{eqnarray}
where the $c_n$ denote the respective expansion coefficients.  Kleinert's
variational perturbation
theory~\cite{HAMPKleinert1} replaces the series (\ref{HAMPEX1}) by\footnote{In
contrast to the standard notation~\cite{HAMPKleinert1} the parameter $q$
in Eq. (\ref{HAMPVPT}) has been chosen in such a way that it coincides
with the critical exponent\index{critical exponent} 
$\Omega$ describing the approach to
scaling in the field theory of critical 
phenomena\index{critical phenomena}~\cite{HAMPVerena}.}
\begin{eqnarray}
\label{HAMPVPT}
f^N ( g , p , q , w ) = \left( \frac{g}{w} \right)^{pq} 
\sum_{n=0}^N c_n w^n \sum_{k=0}^{N-n} (-1)^k
\left( \begin{array}{@{}c} p - 
{\displaystyle \frac{n}{q}} \\ k \end{array} 
\right) \, \left( 1 - \left( \frac{w}{g} \right)^q \right)^k \, .
\end{eqnarray}
By doing so, the newly introduced parameters $N$, $p$, $q$ , $w$ are
specified as follows.  $N$ represents the order of the expansion and
will be increased step by step as far as the knowledge of the
weak-coupling coefficients $c_n$ permits. The parameters $p$ and $q$
determine the asymptotic behaviour of the function $f(g)$ in the
strong-coupling limit\index{strong-coupling limit} 
$g \rightarrow \infty$ according to
\begin{eqnarray}
\label{HAMPSTRONG}
f ( g ) = g^{p q} \sum_{j=0}^{\infty} b_j g^{- jq} \, ,
\end{eqnarray}
where the $b_j$ denote the strong-coupling coefficients.  In quantum
statistics the parameters $p$ and $q$ are usually known, e.g.  from
dimensional arguments, whereas in statistical
field theory they are related to
unknown critical exponents\index{critical exponent}, 
so they have to be determined
self-consistently from the weak-coupling coefficients
$c_n$~\cite{HAMPKleinert2,HAMPKleinert3}.  Finally, as the variational
parameter $w$ is introduced such that the
approximant (\ref{HAMPVPT}) will not depend on it in the limit $N
\rightarrow \infty$, it should be determined according to the principle
of minimal sensitity~\cite{HAMPStevenson}. This principle is, however,  no 
clear-cut mathematical statement and may therefore be implemented 
differently, giving rise to different
versions of Kleinert's variational perturbation theory. Here 
we investigate and compare the versions, which define minimal 
sensitivity 
by the vanishing of a derivative of (\ref{HAMPVPT})
with respect to $w$ to some order $k$. Such
a version is considered to be good, if its results converge well to
the true value in the quantum-statistical case, especially when $p$ is
known, but $q$ is taken to be unknown, since this will anticipate the
situation for the field-theoretical application. Another desirable
feature can be seen in a very smooth dependence on the order $N$,
since this will greatly enhance the possibility to extrapolate 
field-theoretical results to $N=\infty$. 
Finally, the simpler the version's defining
prescription, the more easily it may be generalized to
field-theoretical applications.

To begin with we focus our attention to
the versions of variational perturbation theory
in the strong-coupling limit\index{strong-coupling limit} 
$g \rightarrow \infty$. 
Comparing the Eqs. (\ref{HAMPVPT}) and (\ref{HAMPSTRONG}) gives,
for the leading strong-coupling coefficient $b_0$, the 
expression
\begin{eqnarray}
\label{HAMPBZERO}
b_0^N = \left( -1 \right)^{N} w^{-pq}
\sum_{n=0}^N c_n \left({- w} \right)^n\,
\left( \begin{array}{@{}c} p - 1 - {\displaystyle \frac{n}{q}} \\ N-n 
\end{array} 
\right) \,,
\end{eqnarray}
where the variational parameter $w$ is still to be optimized.
For the ground-state energy of
the anharmonic oscillator\index{anharmonic oscillator} potential
\begin{eqnarray}
\label{HAMPPOT} 
V(x)={\displaystyle \frac{1}{2}}x^2+gx^4
\end{eqnarray}
the leading strong-coupling coefficient has the numerical value~\cite{HAMPCizek}
\begin{equation}
\label{HAMPEXPER} 
b_0=0.667~986~259~155~777~108~270~96...
\end{equation}
and dimensional arguments require $q=2/3$, $p=1/2$.
The weak-coupling coefficients 
$c_n$ are derived in the Appendix (see Tab.~\ref{HAMPTabIV}).
\section{The Highest-Derivative Version}
\begin{figure}[t]
\begin{center}
\setlength{\unitlength}{1cm}
\hspace*{3cm} \begin{picture}(19, 5.8)
\scalebox{.75}[.75]{\includegraphics*{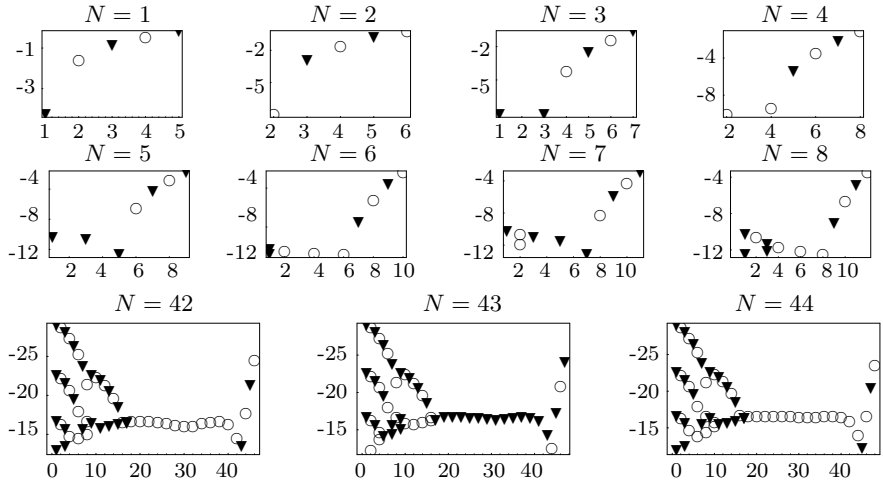}}
\put(-10.6,5.95){\small $N=1$}
\put(-7.6,5.95){\small $N=2$}
\put(-4.55,5.95){\small $N=3$}
\put(-1.6,5.95){\small $N=4$}
\put(-10.6,4.1){\small $N=5$}
\put(-7.6,4.1){\small $N=6$}
\put(-4.55,4.1){\small $N=7$}
\put(-1.6,4.1){\small $N=8$}
\put(-10.2,2.1){\small $N=42$}
\put(-6.1,2.1){\small $N=43$}
\put(-1.95,2.1){\small $N=44$}
\put(-0.7,-0.1){\footnotesize 40}
\put(-1.25,-0.1){\footnotesize 30}
\put(-1.8,-0.1){\footnotesize 20}
\put(-2.35,-0.1){\footnotesize 10}
\put(-2.8,-0.1){\footnotesize 0}
\put(-4.8,-0.1){\footnotesize 40}
\put(-5.3,-0.1){\footnotesize 30}
\put(-5.9,-0.1){\footnotesize 20}
\put(-6.5,-0.1){\footnotesize 10}
\put(-7.0,-0.1){\footnotesize 0}
\put(-8.85,-0.1){\footnotesize 40}
\put(-9.4,-0.1){\footnotesize 30}
\put(-10.0,-0.1){\footnotesize 20}
\put(-10.6,-0.1){\footnotesize 10}
\put(-11.1,-0.1){\footnotesize 0}
\put(-3.35,0.45){\footnotesize -15}
\put(-3.35,0.95){\footnotesize -20}
\put(-3.35,1.45){\footnotesize -25}
\put(-7.5,0.45){\footnotesize -15}
\put(-7.5,0.95){\footnotesize -20}
\put(-7.5,1.45){\footnotesize -25}
\put(-11.6,0.45){\footnotesize -15}
\put(-11.6,0.95){\footnotesize -20}
\put(-11.6,1.45){\footnotesize -25}
\put(-10.85,2.55){\footnotesize 2}
\put(-10.45,2.55){\footnotesize 4}
\put(-9.95,2.55){\footnotesize 6}
\put(-9.5,2.55){\footnotesize 8}
\put(-8,2.55){\footnotesize 2}
\put(-7.55,2.55){\footnotesize 4}
\put(-7.15,2.55){\footnotesize 6}
\put(-6.8,2.55){\footnotesize 8}
\put(-6.5,2.55){\footnotesize 10}
\put(-4.9,2.55){\footnotesize 2}
\put(-4.55,2.55){\footnotesize 4}
\put(-4.15,2.55){\footnotesize 6}
\put(-3.8,2.55){\footnotesize 8}
\put(-3.55,2.55){\footnotesize 10}
\put(-1.75,2.55){\footnotesize 2}
\put(-1.45,2.55){\footnotesize 4}
\put(-1.15,2.55){\footnotesize 6}
\put(-0.8,2.55){\footnotesize 8}
\put(-0.6,2.55){\footnotesize 10}
\put(-2.35,3.8){\footnotesize -4}
\put(-2.35,3.3){\footnotesize -8}
\put(-2.5,2.8){\footnotesize -12}
\put(-5.4,3.8){\footnotesize -4}
\put(-5.4,3.3){\footnotesize -8}
\put(-5.55,2.8){\footnotesize -12}
\put(-8.55,3.8){\footnotesize -4}
\put(-8.55,3.3){\footnotesize -8}
\put(-8.7,2.8){\footnotesize -12}
\put(-11.45,3.8){\footnotesize -4}
\put(-11.45,3.3){\footnotesize -8}
\put(-11.6,2.8){\footnotesize -12}
\put(-11.2,4.4){\footnotesize 1}
\put(-10.75,4.4){\footnotesize 2}
\put(-10.3,4.4){\footnotesize 3}
\put(-9.85,4.4){\footnotesize 4}
\put(-9.4,4.4){\footnotesize 5}
\put(-8.2,4.4){\footnotesize 2}
\put(-7.75,4.4){\footnotesize 3}
\put(-7.3,4.4){\footnotesize 4}
\put(-6.85,4.4){\footnotesize 5}
\put(-6.4,4.4){\footnotesize 6}
\put(-5.15,4.4){\footnotesize 1}
\put(-4.85,4.4){\footnotesize 2}
\put(-4.55,4.4){\footnotesize 3}
\put(-4.25,4.4){\footnotesize 4}
\put(-3.95,4.4){\footnotesize 5}
\put(-3.65,4.4){\footnotesize 6}
\put(-3.35,4.4){\footnotesize 7}
\put(-2.1,4.4){\footnotesize 2}
\put(-1.55,4.4){\footnotesize 4}
\put(-0.95,4.4){\footnotesize 6}
\put(-0.35,4.4){\footnotesize 8}
\put(-2.45,4.9){\footnotesize -8}
\put(-2.45,5.4){\footnotesize -4}
\put(-5.45,5.0){\footnotesize -5}
\put(-5.45,5.5){\footnotesize -2}
\put(-8.45,5.0){\footnotesize -5}
\put(-8.45,5.5){\footnotesize -2}
\put(-11.5,5.0){\footnotesize -3}
\put(-11.5,5.5){\footnotesize -1}
\end{picture}
\end{center}
\caption[FigI]{\label{HAMPFigI} 
The approach of $b_0^N$ to its true value $b_0$ is measured by
the quantity $\log |b_0^N-b_0|$ which
is plotted over $k$, where $k$ appears in 
the condition $\partial^k b^N_0 / \partial w^k=0$ determining the 
variational
parameter $w$. 
Some typical values for the order $N$ have been depicted with
different symbols $\blacktriangledown$ and 
\begin{picture}(4,4)
\put(2,2){\circle{4}} \hspace*{2mm}
\end{picture}
used to plot 
odd and even $k$, respectively. 
Optimal results are 
obtained for $k=1$ or $k=2$, but with no simple rule as to which of the 
two has to be chosen. Taking $k=N$, however, the same quality is obtained 
with no ambiguity of choice. Also for $k=N-2$, the outcome is 
very reasonable. Note that for some of the lower values of 
$k$ there is more than one solution and that
for some $k$, like e.g. for $k=N-1$,
there is no solution at all.}
\end{figure}

Traditionally, the expression (\ref{HAMPBZERO}) for $b_0$
is made stationary by forcing its first or second  
derivative with respect to the variational parameter
$w$ to vanish, depending on whether the order $N$ 
is odd or even. Here we investigate whether some higher derivatives may 
be used 
instead. To this end we consider, for
every order $N$, all derivatives 
$\partial^k b^N_0 / \partial w^k$ with $k=1,\ldots, 
N+4$ and determine all real positive zeros $w$. For each of these zeros, 
the quantity $\log |b_0^N-b_0|$ measures the quality of approximating $b_0$ in
Eq. (\ref{HAMPEXPER}) by  $b^N_0$ in Eq. (\ref{HAMPBZERO})
(see Fig.~\ref{HAMPFigI}). 

Optimal results are obtained for $k=N$. Moreover, the prescription is unique, 
since for $k=N$ there exists only one real positive zero $w$
of $\partial^k b^N_0 / \partial w^k$. It should be 
noted that for $k=N-1$ there never exists a real positive zero. 
Another unique and almost optimal prescription 
would be $k=N-2$ which works well for all $N>2$.
The results for $k=1$  and $k=2$ 
are of comparable quality with the values for $k=N$. 

Although
only one of them has a real positive zero up to $N=5$, 
the equations for $k=1$ and $k=2$ 
tend to have more than one solution for larger $N$, the one 
with the largest value for $w$ always producing the best result. But 
there is 
no simple rule to decide whether $k=1$ or $k=2$ gives the better value. 
The 
initial indication of using $k=1$ for odd and $k=2$ for even orders does 
not 
carry through beyond $N=6$. 

\begin{figure}[t]
\begin{center}
\setlength{\unitlength}{1cm}
\begin{picture}(10,10.2)
\hspace*{0.6cm} \scalebox{.7}[.7]{\includegraphics*{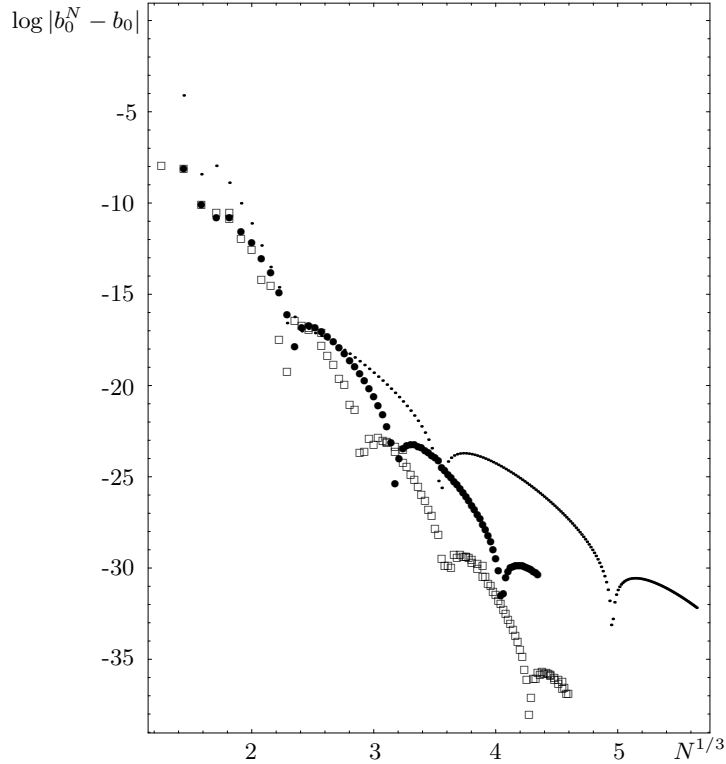}}
\put(-8.1,1.15){\small -35}
\put(-8.1,2.35){\small -30}
\put(-8.1,3.55){\small -25}
\put(-8.1,4.75){\small -20}
\put(-8.1,6.){\small -15}
\put(-8.1,7.2){\small -10}
\put(-7.9,8.4){\small -5}
%
\put(-9.3,9.63){\small $\log |b_0^N-b_0|$}
\put(-0.5,-0.1){\small $N^{1/3}$}
\put(-1.3,-0.1){\small 5}
\put(-2.95,-0.1){\small 4}
\put(-4.55,-0.1){\small 3}
\put(-6.2,-0.1){\small 2}
\end{picture}
\end{center}
\caption[FigII]{\label{HAMPFigII} The error $\log |b_0^N-b_0|$ plotted over 
$N^{1/3}$ as obtained by 
various methods: $\bullet$ for $k=N$, $\cdot$ for a quadratic 
approximation 
to $k=N-2$, and $\Box$ for optimal values from $k=1$ or $k=2$, 
respectively.}
\end{figure}

In Fig.~\ref{HAMPFigII} we compare the results for $k=N$ with 
the optimal branch chosen judiciously from the equations for $k=1$ and 
$k=2$. 
We also show a quadratic approximation to the $k=N-2$ equation where
only the last three terms of the condition 
$\partial^{N-2} b^N_0 / \partial w^{N-2}=0$ are kept,
leaving us with a quadratic equation for the variational parameter
$w$. It can be seen 
that this approximation is of high quality, as was to be expected.

This may be understood as follows: the coefficients of the 
$w$-expansion of 
$b^N_0$ in Eq. (\ref{HAMPBZERO})
alternate in sign  and grow faster than $n!$, which 
is the same for the weak-coupling 
coefficients $c_n$ from which the former 
are derived. Therefore the series (\ref{HAMPBZERO}),
or any derivative thereof will preferably 
become zero, if neighbouring terms in the sum nearly cancel each 
other. 
The failure of such cancellation among some high order  
neighbours cannot easily
be compensated by lower order terms because of the large order behaviour
of the coefficients. Hence the vanishing of the 
derivative for $k=N-2$  
can well be approximated with little loss of accuracy
by choosing $w$ in such a way, that only the three 
highest order terms cancel each other. Thus we 
conclude that $k=N-2$ represents
the highest non-trivial derivative version for 
applications in statistical
field theory, where we usually have $p=0$, so the strong-coupling
coefficient $b_0^N$ in Eq. (\ref{HAMPBZERO}) is a polynomial in $w$.
It becomes the simpler the higher derivatives are being used.
\pagebreak
\vspace*{1.5cm}
\begin{figure}[h]
\begin{center}
\setlength{\unitlength}{1cm}
\begin{picture}(12, 5.8)
\hspace*{2.3cm}\scalebox{.6}[.6]{\includegraphics*{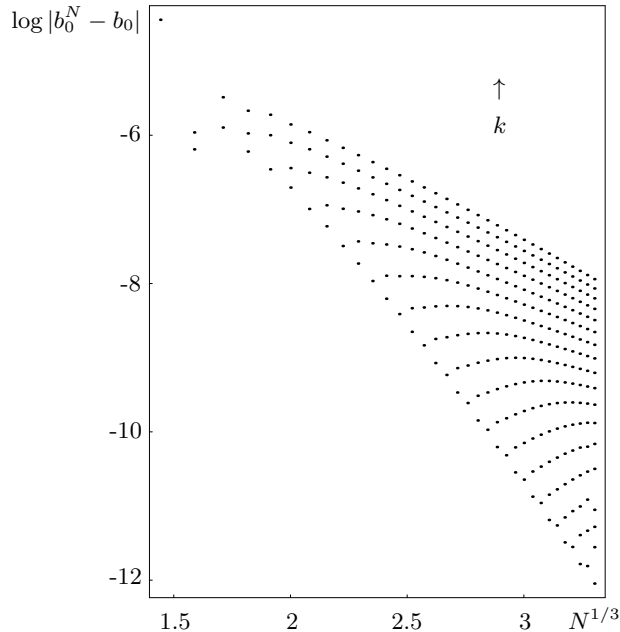}}
\put(-7.9,7.9){\small $\log |b^N_0-b_0|$}
\put(-0.5,-0.1){\small $N^{1/3}$}
\put(-1.5,6.5){\small $k$}
\put(-1.5,7){\small $\uparrow$}
\put(-1.15,-0.1){\small 3}
\put(-2.85,-0.1){\small 2.5}
\put(-4.25,-0.1){\small 2}
\put(-5.93,-0.1){\small 1.5}
\put(-6.6,0.5){\small -12}
\put(-6.6,2.45){\small -10}
\put(-6.45,4.4){\small -8}
\put(-6.45,6.4){\small -6}
\end{picture}
\end{center}
\caption[FigIII]{\label{HAMPFigIII} For every $N$ several $q$-values 
are obtained by setting the 
$k$th derivative of the 
strong-coupling limit\index{strong-coupling limit} 
of $F (g)=g f{}'(g)/f(g)$ 
with respect to the variational parameter $w$
equal to zero. For each of these self-consistent
values of $q$ the error
$\log|b_0^N-b_0|$ is plotted over $N^{1/3}$. Smaller $k$ give better 
results, but the highest $k$-values up to $k=N-1$ are also tolerable.}
\end{figure}
\section{Self-Consistent Determination of the Parameter $q$}
In this section we evaluate again the leading coefficient $b_0$  
in the strong-coupling expansion for the ground-state energy of the
anharmonic oscillator\index{anharmonic oscillator}. 
But this time the parameter $q$ is not fixed to its 
proper value $q=2/3$. Instead we draw this information from the same 
weak-coupling coefficients $c_n$, whereas  $p$ is set to its known 
value $p=1/2$. In order to determine $q$ we need an 
additional equation. This usually results from a self-consistent 
reasoning by evaluating the logarithmic derivative of the strong-coupling 
series (\ref{HAMPSTRONG}) of $f (g)$
in the strong-coupling limit\index{strong-coupling limit} 
$g \rightarrow \infty$~\cite{HAMPKleinert2,HAMPKleinert3}:
\begin{equation}
\label{HAMPLOGDIFF}
\lim_{g\to \infty}\frac{gf {}'(g)}{f (g)}=
\lim_{g\to \infty}
\frac{\displaystyle g^{pq}\sum_{j=0}^{\infty}b_j q(p-j)
g^{-jq}}{\displaystyle g^{pq}\sum_{j=0}^{\infty}b_j 
g^{-jq}}=pq=\frac{q}{2} \,.
\end{equation}
Defining $F(g):= g f'(g) / f(g)$, the weak-coupling 
coefficients $\gamma_n$ of $F(g)$ are determined from 
the corresponding $c_n$ of $f ( g)$
in Eq. (\ref{HAMPEX1}).  
The strong-coupling limit\index{strong-coupling limit} 
of $F(g)$ can then be constructed with
the help of variational perturbation theory by using 
$\gamma_n$ 
instead of $c_n$ in quite an analogous way as before. Since $F(g)$ 
tends to the constant value $q/2$ in the 
strong-coupling limit\index{strong-coupling limit}
$g \rightarrow \infty$, its parameter $p$ 
has to be zero, whereas $q$ must have the same 
value as before. Thus, from the Eqs. (\ref{HAMPBZERO}) and (\ref{HAMPLOGDIFF}),
we obtain for the strong-coupling limit\index{strong-coupling limit} of $F(g)$
\begin{equation}
\label{HAMPQQQQ}
(-1)^N \sum_{n=0}^{N} \gamma_n (-w)^n
\left( \begin{array}{@{}c} -1 - {\displaystyle \frac{n}{q}} \\ 
N-n \end{array} \right) = \frac{q}{2} \,,
\end{equation}
where  we determine the variational parameter $w$ on the left-hand side
by demanding that the $k$th derivative with 
respect to $w$ must vanish, $k$ being chosen appropriately. This set of 
equations can be rearranged into two polynomials of $q$ and $w$ which 
have to vanish simultaneously. The complete set of complex zeros becomes 
legion even for moderate orders $N$, their number growing with $N^2$. 
Constraining ourselves to real and positive solutions for $q$ in the 
expected 
range  $0.5<q<0.9$, we find, in contrast to the previous case,  a very 
regular behaviour. 
Exactly one unique solution exists for all $N$ up to 
$N=36$ 
and for all $k$ if $N+k$ is an odd integer, whereas there 
is no solution within 
this range if $N+k$ is even. The solutions for different $k$ approach the 
value $q=2/3$ closer for smaller $k$. As $k$ increases,
the loss of accuracy, however,  is tolerable, 
as can be seen from Fig.~\ref{HAMPFigIII}, 
where again the error $\log |b_0^N-b_0|$ is plotted over $N^{1/3}$, the 
approximation
$b^N_0$  being obtained as before with the highest derivative version, but 
using the self-consistently determined $q$
instead of the exact 
$q=2/3$. Some loss of accuracy is the price 
to pay, if we want to make one of both equations linear by choosing  
$k=N-1$, which is now the highest non-trivial derivative. Notice that the 
penalty is relatively low for small orders $N$, which in field-theoretical 
models are the only ones available at present. 

\section{Critical Exponent $\boldmath{\alpha}$ 
for Liquid Helium}\index{critical exponent}\index{helium}
After having analysed the highest derivative versions of variational 
perturbation theory by the example of the ground-state energy of the
anharmonic oscillator\index{anharmonic oscillator}, 
we apply this method to liquid 
${}^4$He. 
In particular, we consider
its superfluid state near the transition point $T_c$, 
in order to calculate the critical exponent\index{critical exponent} 
$\alpha$ governing the power 
behaviour $|T_c-T|^{-\alpha}$ 
of the specific heat\index{specific heat}. 
Within the framework
of the $\phi^4$-theory of critical phenomena\index{critical phenomena}, 
the superfluid phase 
transition
of ${}^4$He is 
described by a complex order parameter field $\phi$ whose 
bare energy functional is of the Ginzburg-Landau 
type and reads in $d=3$ dimensions~\cite{HAMPVerena}: 
\begin{equation}
\label{HAMPEFUNC}
E\left[\phi_B \right]=\int d^3x \left\{
\frac{1}{2}\left[\partial\phi_B(x)\right]^2+\frac{1}{2}\,m_B^2\phi^2_B(x)+
\frac{\pi}{5} \,
g_B \left[\phi_B^2(x)\right]^2 \right\} \,.
\end{equation}
By calculating the Feynman diagrams, one encounters divergencies which
are removed by analytic regularization~\cite{HAMPParisi}. A subsequent 
renormalization leads to renormalized values of mass, coupling
constant and field which are 
related to the bare input quantities by the
respective renormalization constants $Z_m$, $Z_g$ and $Z_{\phi}$:
\begin{eqnarray}
m_B^2 = m^2 Z_m Z_{\phi}^{-1} \, , \hspace*{0.5cm}
g_B = g Z_g Z_{\phi}^{-2} \, , \hspace*{0.5cm}
\phi_B = \phi Z_{\phi}^{1/2} \, .
\end{eqnarray}
In the literature, one finds expansions for these renormalization
constants and for certain logarithmic derivatives, the so-called
renormalization group functions, up to six and partly up so seven loops~\cite{HAMPNickel1,HAMPSokolov,HAMPNickel2}. All these expansions can be written
in terms of the dimensionless bare coupling constant
$\bar{g}_B = g_B / m$.
For instance, one obtains, for the dimensionless
renormalized coupling constant $\bar{g} = g / m$, the series
\begin{eqnarray}
\label{HAMPG}
\bar{g} ( \bar{g}_B ) = \sum_{n=0}^{\infty} \kappa_n \bar{g}^n_B \, .
\end{eqnarray}
The logarithmic derivative of the square mass ratio
\begin{eqnarray}
\eta_m ( \bar{g}_B ) = - \frac{d \log {\displaystyle
\frac{m^2}{m_B^2}}}{d \log \bar{g}_B}
\end{eqnarray}
reads correspondingly
\begin{eqnarray}
\label{HAMPETA}
\eta_m (\bar{g}_B ) = \sum_{n=0}^{\infty} \lambda_n \bar{g}^n_B \, .
\end{eqnarray}
Its weak-coupling coefficients
$\kappa_n$ and $\lambda_n$  have been listed in Tab.~\ref{HAMPTabI},
respectively.

In the field-theoretic description of second-order phase transitions, the
renormalized square mass $m^2$ vanishes near the transition point $T_c$. 
Thus the approach to scaling corresponds to the 
strong-coupling limit\index{strong-coupling limit}
of the dimensionless bare coupling constant $\bar{g}_B$~\cite{HAMPVerena,HAMPKleinert2,HAMPKleinert3,HAMPKleinert4,HAMPKleinert5,HAMPKleinert6}.
Expecting to obtain finite results for both (\ref{HAMPG}) and (\ref{HAMPETA}),
when the dimensionless bare coupling constant $\bar{g}_B$ tends to 
infinity, the strong-coupling expansions for $\bar{g} (\bar{g}_B)$
and $\eta_m (\bar{g}_B)$ should read
\begin{eqnarray}
\bar{g} (\bar{g}_B) = \bar{g} + \sum_{j=1}^{\infty} \beta_j 
\bar{g}_B^{-j \Omega} 
\, , \label{HAMPGS} \\
\eta_m ( \bar{g}_B) = \eta_m + \sum_{j=1}^{\infty} \gamma_j 
\bar{g}_B^{-j\Omega} 
\, . \label{HAMPETAS}
\end{eqnarray} 
\pagebreak
\begin{table}[t]
\caption[TabI]{\label{HAMPTabI} Weak-coupling 
coefficients\index{weak-coupling!coefficients}\index{coefficients!weak-coupling} 
for $\bar{g}(\bar{g}_B)$ and $\eta_m(\bar{g}_B)$ from 6- 
and 7-loop perturbation theory~\cite{HAMPNickel1,HAMPSokolov,HAMPNickel2}
(above the tripple line) and from 
extrapolations based on instanton calculations~\cite{HAMPKleinert4,HAMPParisi} (below the tripple line).}
\begin{center}
\begin{minipage}{9.5cm}
\begin{tabular}{|c|r|r|}\hline
\multicolumn{3}{|c|}{\bf Weak-Coupling Coefficients} \\ \hline
  & \multicolumn{1}{c|}{ $\eta_m =\sum \lambda_n \bar{g}_B^n $} & 
\multicolumn{1}{c|}{ $\bar{g}=\sum \kappa_n \bar{g}_B^n $} \\ \hline
$n$ & \multicolumn{1}{c|}{ $\lambda_n$} & \multicolumn{1}{c|}{ 
$\kappa_n $} 
\\ \hline \hline
$0 $ & $ 0 $ & $ 0 $ \\ \hline
$1 $ & $ 4 $ & $ 1 $ \\ \hline
$2 $ & $ -46.814\,814\,814\,814\,82 $ & $ -10 $ \\ \hline
$3 $ & $ 667.389\,693\,519\,318\,6 $ & $ 120.148\,148\,148\,148\,1 $ \\ 
\hline
$4 $ & $ -10\,792.618\,387\,448\,09 $ & $ -1\,642.256\,264\,617\,284 $ \\ 
\hline
$5 $ & $ 191\,274.332\,379\,005\,1 $ & $ 24\,816.045\,615\,887\,86 $ \\ 
\hline
$6 $ & $ -3\,644\,347.117\,315\,811$ & $ -407\,363.539\,559\,348 $ \\ 
\hline
$7 $ & $ 7\,378\,080.984\,900\,002 $ &  \\ \hline \hline 
$7 $ &  & $ 7.180\,326\,000\,784\,143 \times E^{6} $ \\ \hline
$8 $ & $ -1.575\,312\,899\,817\,985 \times E^{9} $ & $ 
-1.347\,981\,388\,551\,665 \times E^{8} $ \\ \hline
$9 $ & $ 3.529\,433\,822\,947\,775 \times E^{10} $ & $ 
2.679\,259\,494\,762\,891 \times E^{9} $ \\ \hline
$10 $ & $ -8.269\,004\,838\,427\,051 \times E^{11} $ & $ 
-5.612\,792\,935\,327\,37 \times E^{10} $ \\ \hline
$11 $ & $ 2.020\,940\,372\,700\,199 \times E^{13} $ & $ 
1.234\,985\,211\,831\,956 \times E^{12} $ \\ \hline
$12 $ & $ -5.143\,391\,961\,273\,287 \times E^{14} $ & $ 
-2.846\,297\,378\,479\,211 \times E^{13} $ \\ \hline
$13 $ & $ 1.360\,628\,154\,286\,311 \times E^{16} $ & $ 
6.837\,009\,909\,070\,767 \times E^{14} $ \\ \hline
$14 $ & $ -3.734\,409\,500\,947\,708 \times E^{17} $ & $ 
-1.699\,763\,366\,416\,082 \times E^{16} $ \\ \hline
$15 $ & $ 1.062\,320\,774\,475\,501 \times E^{19} $ & $ 
4.354\,392\,675\,760\,15 \times E^{17} $ \\ \hline
$16 $ & $ -3.132\,670\,751\,194\,59 \times E^{20} $ & $ 
-1.151\,470\,001\,163\,625 \times E^{19} $ \\ \hline
$17 $ & $ 9.584\,565\,077\,380\,565 \times E^{21} $ & $ 
3.166\,449\,929\,095\,059 \times E^{20} $ \\ \hline
$18 $ & $ -3.044\,620\,939\,704\,96 \times E^{23} $ & $ 
-9.135\,728\,488\,698\,395 \times E^{21} $ \\ \hline
$19 $ & $ 1.004\,119\,260\,969\,699 \times E^{25} $ & $ 
2.780\,684\,275\,579\,437 \times E^{23} $ \\ \hline
$20 $ & $ -3.435\,379\,417\,189\,77 \times E^{26} $ & $ 
-8.925\,916\,735\,892\,876 \times E^{24} $ \\ \hline
$21 $ & $ 1.217\,685\,465\,986\,233 \times E^{28} $ & $ 
3.006\,538\,451\,329\,012 \times E^{26} $ \\ \hline
$22 $ & $ -4.465\,282\,353\,357\,697 \times E^{29} $ & $ 
-1.055\,6247\,984\,629\,23 \times E^{28} $ \\ \hline
$23 $ & $ 1.691\,830\,517\,012\,908 \times E^{31} $ & $ 
3.841\,496\,075\,948\,586 \times E^{29} $ \\ \hline
$24 $ & $ -6.616\,032\,677\,503\,413 \times E^{32} $ & $ 
-1.443\,458\,944\,165\,208 \times E^{31} $ \\ \hline 
\end{tabular}
\end{minipage}
\end{center}
\end{table}
\vspace*{-1cm}
\begin{table}[t]
\caption[TabII]{\label{HAMPTabII} The critical exponent\index{critical!exponent} 
$\Omega$  
as obtained from the 
strong-coupling limit\index{strong-coupling!limit}\index{limit!strong-coupling}  
of the first or the second logarithmic derivatives\index{logarithmic!derivative}\index{derivative!logarithmic} of the 
weak-coupling series\index{weak-coupling!expansion}\index{expansion!weak-coupling} 
for $\bar{g} ( \bar{g}_B) $ and $\eta_m (\bar{g}_B)$, respectively.}
\begin{center}
\begin{minipage}{7.5cm}
\begin{tabular}{|c|r|r|r|r|}\hline 
\multicolumn{5}{|c|}{\bf $\Omega$-Values} \\ \hline
 $N$ & \multicolumn{1}{c|}{ $\Omega(\bar{g},1)$} & 
\multicolumn{1}{c|}{$\Omega(\bar{g},2)$} & 
\multicolumn{1}{c|}{ $\Omega(\eta_m,1)$} & 
\multicolumn{1}{c|}{ $\Omega(\eta_m,2)$}  \\ \hline \hline
$2 $ & $ 0.730\,495 $ & $ 0.735\,397 $ & $ 0.715\,930 $ & $ 
0.721\,303$ \\ \hline
$3 $ & $ 0.751\,627 $ & $ 0.751\,166 $ & $ 0.714\,270 $ & $ 
0.712\,744$ \\ \hline
$4 $ & $ 0.757\,596 $ & $ 0.757\,762 $ & $ 0.703\,544 $ & $ 
0.700\,880$ \\ \hline
$5 $ & $ 0.763\,865 $ & $ 0.763\,975 $ & $ 0.705\,086 $ & $ 
0.704\,576$ \\ \hline
$6 $ & $  $ & $  $ & $ 0.714\,586 $ & $ 0.714\,540$ \\ \hline 
\end{tabular}
\end{minipage}
\end{center}
\end{table}
\vspace*{-1cm}
\begin{table}[t]
\caption[TabIII]{\label{HAMPTabIII} The critical 
exponent\index{critical!exponent} $\alpha$ 
for the specific heat\index{specific heat} of 
liquid helium\index{helium}, 
derived from the perturbation expansion for $\eta_m$ using 
the highest-derivative version of variational perturbation 
theory\index{variational!perturbation theory}\index{theory!variational perturbation}.}
\begin{center}
\begin{minipage}{3.5cm}
\begin{tabular}{|c|r|r|}\hline 
\multicolumn{3}{|c|}{\bf $\alpha$-Values} \\ \hline
 $N$ & \multicolumn{1}{c|}{ $\eta_m$} & \multicolumn{1}{c|}{ $\alpha$}  
\\ \hline \hline
$2 $ & $ 0.490\,834 $ & $ +0.0121 $ \\ \hline
$3 $ & $ 0.513\,786 $ & $ -0.0185 $ \\ \hline
$4 $ & $ 0.522\,480 $ & $ -0.0304 $ \\ \hline
$5 $ & $ 0.522\,651 $ & $ -0.0307 $ \\ \hline
$6 $ & $ 0.519\,592 $ & $ -0.0265 $ \\ \hline 
\end{tabular}
\end{minipage}
\end{center}
\end{table}
Note that strong-coupling expansions like (\ref{HAMPGS}) and (\ref{HAMPETAS}) are
governed by one and the same critical exponent\index{critical exponent} 
$\Omega$~\cite{HAMPVerena}, a common feature of 
second-order phase transitions.
Thus from either of these series $\Omega$  may be extracted 
in various ways, 
taking logarithmic derivatives with respect to 
the dimensionless bare coupling constant $\bar{g}_B$. The 
strong-coupling behaviour (\ref{HAMPGS}) of the dimensionless
renormalized coupling constant $\bar{g}(\bar{g}_B)$
implies, for instance,
\begin{eqnarray}
\label{HAMPLIMITS}
\lim_{\bar{g}_B \rightarrow \infty} \frac{d \log \bar{g}(\bar{g}_B)}{d
\log \bar{g}_B} &=&0 \, ,\\
\lim_{\bar{g}_B \rightarrow \infty} \frac{d \log \bar{g}'(\bar{g}_B) }{d 
\log \bar{g}_B}&=& - \Omega-1 \, , \label{HAMPLIMITS2}
\end{eqnarray}
with corresponding results for $\eta_m(\bar{g}_B)$, 
as can easily been seen from Eq. (\ref{HAMPETAS}). Forcing 
these relations upon the weak-coupling expansions as well, 
we obtain equations to which the 
strong-coupling limit\index{strong-coupling limit} 
of variational 
perturbation theory can be applied. We use its highest derivative 
version as explained above
and find good agreement between the $\Omega_N$-values from each of the 
four equations (see Tab.~\ref{HAMPTabII}).

In fact they are supposed to approach the same limit as 
$N\rightarrow\infty$.
We notice, that the members of each pair obtained from either the 
$\eta_m$- 
or the $\bar{g}$- expansion are almost the same, whereas the different 
pairs have 
not yet converged so well for $N\leq 6$. 
Even though the data from the $\bar{g}$-expansion look more 
promising 
because of their smoother behaviour, we still use the results $\Omega_N$ 
from the $\eta_m$-expansion alone to calculate the 
critical exponent\index{critical exponent}
$\alpha$, and do not use 
any information from the $\bar{g}$-expansion at this stage. 
We evaluate the critical exponent\index{critical exponent}
$\eta_m$ from
\begin{eqnarray}
\label{HAMPETAM}
\eta_m^N = (-1)^N
\sum_{n=0}^N \lambda_n (-w_N)^n \left( \begin{array}{c}
-1 - {\displaystyle \frac{n}{\Omega_N}} \\ N-n \end{array}
\right) \, ,
\end{eqnarray}
with the variational parameter $w_N$ fixed by
\begin{eqnarray}
w_N&=&-\frac{\lambda_{N-1}(N-1+\Omega_N)}{\lambda_N N \Omega_N} \,.
\end{eqnarray}
Here the $\Omega_N$ are supplied from the 
$\Omega(\eta_m,1)$-column 
of Tab.~\ref{HAMPTabII}.
The corresponding results are shown in Tab.~\ref{HAMPTabIII} along with the 
corresponding values for the critical exponent\index{critical exponent} 
$\alpha$ of the specific heat\index{specific heat}
which follows from 
\begin{equation}
\label{HAMPGETA2}
\alpha=\frac{1-2\eta_m}{\eta_m-2} \,.
\end{equation}
These are values quite close to the 
experimental data $\alpha_{\rm exp}=-0.01056\pm0.00038$~\cite{HAMPLipa}, 
thus giving 
support for the method used.

\begin{figure}[t]
\begin{center}
\setlength{\unitlength}{1cm}
\hspace*{3cm} \begin{picture}(18,6)
\hspace*{0.6cm} \scalebox{.6}[.6]{\includegraphics*{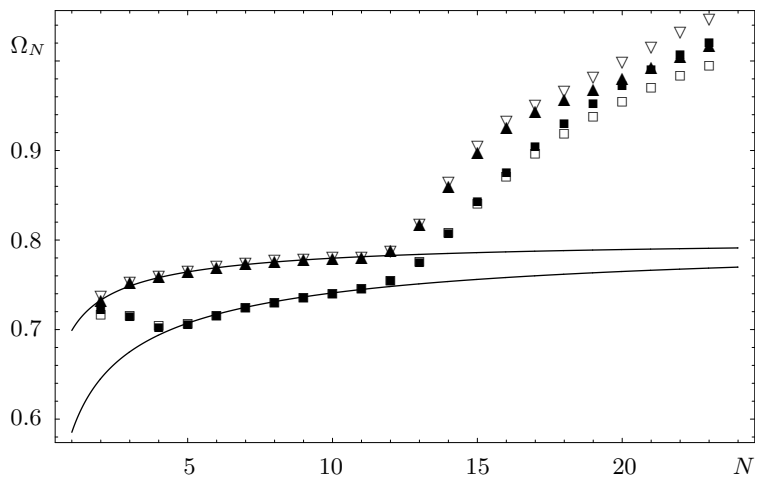}}
\put(-9.9,5.5){\small $\Omega_N$}
\put(-0.3,-0.1){\small $N$}
\put(-1.9,-0.1){\small 20}
\put(-3.85,-0.1){\small 15}
\put(-5.8,-0.1){\small 10}
\put(-7.6,-0.1){\small 5}
\put(-9.9,0.55){\small 0.6}
\put(-9.9,1.73){\small 0.7}
\put(-9.9,2.9){\small 0.8}
\put(-9.9,4.1){\small 0.9}
%
\end{picture}
\end{center}
\caption[FigIV]{\label{HAMPFigIV} The critical 
exponent\index{critical exponent} $\Omega$ 
derived from the 
strong-coupling limit\index{strong-coupling limit} 
of the first and the second
logarithmic 
derivative of the weak-coupling expansion of $\bar{g}(\bar{g}_B)$ 
($\blacktriangle$, $\triangledown$) and of $\eta_m(\bar{g}_B)$ 
($\blacksquare$, $\square$), respectively. 
The solid lines are extrapolations 
with $\Omega_\infty=0.8$ fitted to the $\bar{g}$- and $\eta_m$-points in the 
reliable interval $3<N<12$, respectively.}
\end{figure}
\section{Improvement by Extrapolation}
\begin{figure}[t]
\begin{center}
\setlength{\unitlength}{1cm}
\hspace*{3cm} \begin{picture}(18, 5.5)
\hspace*{0.6cm}\scalebox{.6}[.6]{\includegraphics*{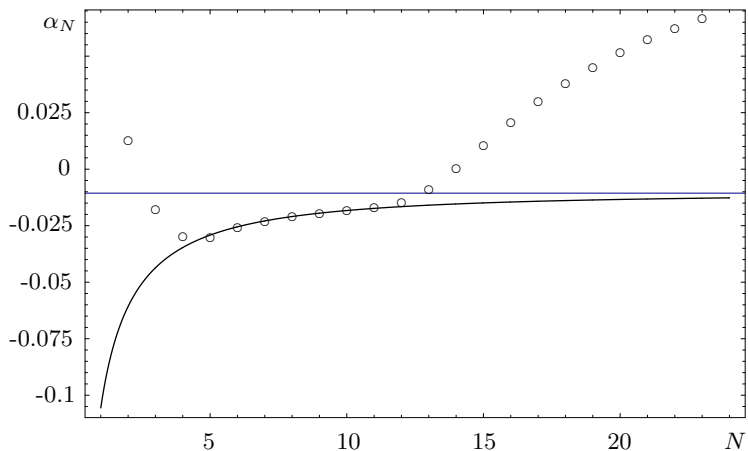}}
\put(-0.3,-0.1){\small $N$}
\put(-9.35,5.5){\small $\alpha_N$}
\put(-1.85,-0.1){\small 20}
\put(-3.65,-0.1){\small 15}
\put(-5.45,-0.1){\small 10}
\put(-7.23,-0.1){\small 5}
\put(-9.45,0.53){\small -0.1}
\put(-9.8,1.27){\small -0.075}
\put(-9.65,2.03){\small -0.05}
\put(-9.8,2.8){\small -0.025}
\put(-9.15,3.55){\small 0}
\put(-9.7,4.3){\small 0.025}
%
\end{picture}
\end{center}
\caption[FigV]{\label{HAMPFigV} The critical exponent\index{critical exponent} 
$\alpha$ 
of the specific heat\index{specific heat} 
of liquid ${}^4$He
near the superfluid phase transition plotted against the order $N$. 
The straight horizontal line represents the experimental value of 
$\alpha_{\rm exp}=-0.0106$. The curved line is an extrapolation of the results 
from the range of  $3<N<12$.}
\end{figure}

Higher order perturbation coefficients for 
$\bar{g}(\bar{g}_B)$ and $\eta_m(\bar{g}_B)$ are 
known approximately (see Tab.~\ref{HAMPTabI}). They have been derived 
from instanton calculations, which were fitted to the weak-coupling data~\cite{HAMPKleinert4,HAMPParisi}. 
Extending the previous calculation on this basis up to 
the order of $N=24$, good convergence of all four different sets of the 
$\Omega_N$ is found. We notice, however, a 
kink at about $N=12$ in Fig.~\ref{HAMPFigIV}, 
which strongly suggests, that the extrapolation is 
no longer reliable beyond this point. 
For large $N$ the results
$\Omega_N$ have the asymptotic form~\cite{HAMPVerena,HAMPKleinert2,HAMPKleinert3,HAMPKleinert4,HAMPKleinert5,HAMPKleinert6}
\begin{equation}
\label{HAMPINFTY}
\Omega_N=\Omega_\infty-a\,\exp \left(-b\,N^{1-\Omega_\infty}\right)\,,
\end{equation}
where the constants $a$ and $b$ are different for the $\eta_m$- and the 
$\bar{g}$- 
expansion. Unfortunately, the fit is extremely insensitive to the value of 
$\Omega_\infty$: all values in a broad interval around $\Omega_\infty=0.8$ 
would fit the data up to $N=12$ from 
both series with very much the same quality 
(see Fig.~\ref{HAMPFigIV}). 
In Fig.~\ref{HAMPFigV} the resulting $\alpha$-values are plotted against 
the asymptotic form
\begin{equation}
\label{HAMPALF}
\alpha_N=\alpha_\infty-c\,\exp \left(-d\,N^{1-\Omega_\infty}\right)\,,
\end{equation}
for $\alpha_\infty=-0.0106$ and $\Omega_\infty=0.8$. Up to the kink at 
$N=12$, the data seem to approach the experimental number excellently.
\begin{table}[t]
\caption[comparison]{\label{HAMPcomparison} Comparison of quality for the 
highest-derivative 
version and for the traditional method, based on convergence of the 
critical exponent\index{critical!exponent} 
$\Omega$  obtained in two different ways from two 
different sets of data. }
\begin{center}
\begin{minipage}{9cm}
\begin{tabular}{|c|c|r|r|r|}\hline 
\multicolumn{5}{|c|}{\bf $\sigma_\Omega$} \\ \hline
data & method & \multicolumn{1}{c|}{$2 \leq N \leq 6$} & 
\multicolumn{1}{c|}{ $7 \leq N \leq 12$} & \multicolumn{1}{c|}{ $13 
\leq N \leq 24$}  \\ \hline \hline
$\bar{g}(\bar{g}_B)$ & high & $ 0.002\,159 $ & $ 0.000\,570 $ & $ 0.017\,682 
$\\ \hline
$\bar{g}(\bar{g}_B)$ & low  & $ 0.002\,245 $ & $ 0.000\,522 $ & $ 0.000\,274 
$\\ \hline \hline
$\eta_m(\bar{g}_B)$&high & $ 0.002\,749 $ & $ 0.000\,258 $ & $ 0.016\,251 $\\ 
\hline
$\eta_m(\bar{g}_B)$ &low & $ 0.008\,214 $ & $ 0.008\,121 $ & $ 0.005\,915 $\\ 
\hline \hline
both          & high & $ 0.046\,663 $ & $ 0.040\,836 $ & $ 0.037\,912 $\\ 
\hline
both          & low  & $ 0.046\,549 $ & $ 0.057\,049 $ & $ 0.055\,477 $\\ 
\hline 
\end{tabular}
\end{minipage}
\end{center}
\end{table} 
\section{Quality Check for the Highest Derivative Version}
The four different methods, discussed before to extract the critical 
exponent\index{critical exponent} 
$\Omega$ from the data, should all give results which converge 
to the same limit as $N\rightarrow\infty$. 
We show now that the highest derivative 
version reveals a better convergence than the traditional method which 
uses the first or second derivatives instead. The details are 
reported in Tab.~\ref{HAMPcomparison}, where the variance 
$\sigma_\Omega$ between the values $\Omega^{(1)}_N$, $\Omega^{(2)}_N$
from different methods is listed for specific ranges of $N$:
\begin{equation}
\label{HAMPVARIANCE}
\sigma_\Omega=\sqrt{\frac{1}{N_1-N_0+1}\sum_{N=N_0}^{N_1} 
(\Omega^{(1)}_N-\Omega^{(2)}_N)^2}\,.
\end{equation}
In the first row of Tab.~\ref{HAMPcomparison} we list the $\Omega$-values 
extracted from the weak-coupling expansion (\ref{HAMPG})
of $\bar{g}(\bar{g}_B)$ according to 
the methods (\ref{HAMPLIMITS}), (\ref{HAMPLIMITS2}) using 
the highest derivative version. The same quantity evaluated in the 
traditional way, i.e. using the first derivative for even $N$ and the 
second derivative for odd $N$, is shown, for comparison, in the second row. 
It can be seen that both methods are of the same quality in the 
range of $2 \leq N \leq 6$, for which the calculation is based on 
the weak-coupling 
coefficients from perturbation theory, and also in the range of 
$7 \leq N \leq 12$, based on extrapolated values. Only for $N>12$, 
where the extrapolation is no more reliable, anyhow, the 
highest derivative version seems to be worse than the one obtained by the
traditional 
method. However, in order to obtain these apparently better results using 
low order derivatives, the rules of the game had to be changed somewhat, 
taking the second derivative throughout for even and odd order $N$, as 
soon as $N$ becomes larger than $12$. 

The next two rows show the corresponding comparison based on the 
weak-coupling expansion (\ref{HAMPETA}) of
$\eta_m(\bar{g}_B)$. Here it turns out that the 
highest derivative version gives results of significantly better quality 
than the traditional method, up to $N=12$.  

Finally, in the last two rows the deviations of the $\Omega$-values from 
the weak-coupling expansions (\ref{HAMPG}) of $\bar{g}(\bar{g}_B)$ 
and (\ref{HAMPETA}) of $\eta_m(\bar{g}_B)$, extracted according to 
(\ref{HAMPLIMITS}) are shown in comparison for the highest derivative 
version (above) and the traditional method (below). It can be seen, 
that both methods are more or less comparable in quality, the highest 
derivative version giving slightly better results.

\section{Appendix}
The perturbative solution of the time-independent Schrödinger equation 
\begin{eqnarray}
\label{HAMPSCHR}
H \Psi_m :=(H_0+g\,V)\Psi_m = E_m\Psi_m
\end{eqnarray}
usually leads to infinite sums for the expansion coefficients of the 
energy 
eigenvalues $E_m(g)$ and the state vectors $\Psi_m(g)$. But if the 
perturbing potential $V$ happens to be band-diagonal in the basis of the
eigenvectors of the unperturbed Hamiltonian
$H_0$, then all these sums become finite and the eigenvalues 
$E_m(g)$  can be determined recursively to all orders. 

Here we assume that the spectrum ${\epsilon_0^{(0)},\,\epsilon_0^{(1)},\,
\epsilon_0^{(2)},\dots,\epsilon_0^{(m)},\dots}$ of 
the unperturbed Hamiltonian $H_0$ is non-degenerate and 
denote the corresponding eigenvectors by 
$|0\rangle, |1\rangle, |2\rangle,\dots,|m\rangle,
\dots$ Following the usual procedure, the 
energy eigenvalues $E_m(g)$ and the state vectors $\Psi_m(g)$ are
expanded in powers of $g$:
\begin{eqnarray}
\label{HAMPSCHR_E}
\Psi_m(g)&=& \sum_{n,i} \gamma_{n,i}^{(m)} g^i \alpha_n |n \rangle \, , \\
E_m(g)&=&\sum_j \epsilon_j^{(m)}g^j\,,
\label{HAMPSCHR_W}
\end{eqnarray}
where the $\alpha_n$ are inserted into the definition, for later 
computational convenience.  They will only show up in the coefficients of  
the state vector $\Psi_m$, but not in the expansion of the energy 
eigenvalues $E_m$ which we are looking for. Without loss of generality 
the 
normalisation of the state vectors $\Psi_m$
is chosen such that $\langle \Psi_m|m \rangle =\alpha_m$ to 
all orders so that we have 
\begin{equation}
\label{HAMPGAMMA}
\gamma_{m,i}^{(m)}=\delta_{i,0} \qquad \gamma_{k,0}^{(m)}=\delta_{k,m}\,.
\end{equation}
Inserting the ansatz (\ref{HAMPSCHR_E}), (\ref{HAMPSCHR_W})
into the Schrödinger equation (\ref{HAMPSCHR}), projecting the result 
onto the base vector $\langle k|\alpha_k$ 
and extracting the coefficient of $g^j$, 
we obtain the relation:
\begin{equation}
\label{HAMPREC1}
\gamma_{k,i}^{(m)}\epsilon_0^{(k)}+\sum_n 
\frac{\alpha_n}{\alpha_k}V_{k,n}\,
\gamma_{n,i-1}^{(m)}=\sum_j\epsilon_j^{(m)}\gamma_{k,i-j}^{(m)}\,,
\end{equation}
with the matrix elements 
\begin{eqnarray}
\label{HAMPMA1}
V_{k,n} := \langle k | V | n \rangle \, .
\end{eqnarray}
For $i=0$ this equation is identically satisfied. For $i\ne 0$ we get the 
following two relations, one for $k=m$ and the other one for $k\ne m$:
\begin{eqnarray}
\label{HAMPEQ1_2}
\epsilon_i^{(m)} &=& \sum_n \gamma_{m+n,i-1}^{(m)} W_{m,n} \, ,  \\
\gamma_{k,i}^{(m)} &=& \frac{1}{\epsilon_0^{(k)}-\epsilon_0^{(m)}} 
\left[\sum_{j=1}^{i-1} \epsilon_j^{(m)}\gamma_{k,i-j}^{(m)}-\sum_n 
\gamma_{k+n,i-1}^{(m)}W_{k,n}\right]\,.
\end{eqnarray}
In these expressions all sums are finite if the potential $V$ is 
band-diagonal such that the augmented matrix elements 
\begin{eqnarray}
\label{HAMPMA2}
W_{m,n}:=V_{m,m+n}\frac{\alpha_{m+n}}{\alpha_m} 
\end{eqnarray}
are different from zero only for $-d<n<d$, where $d$ is some finite 
number. 
For any quantum number $m$ these relations may now be applied recursively 
with respect to the order, leading to 
$\epsilon_1^{(m)},\, \gamma_{k,1}^{(m)},\, \epsilon_2^{(m)},\, 
\gamma_{k,2}^{(m)},\dots,\epsilon_r^{(m)}, \gamma_{k,r}^{(m)},\dots$ 
in turn.
Notice that for given $m$ and $r$ only a finite number of the 
$\gamma_{k,r}^{(m)}$ is non-zero.

Considering in particular the 
anharmonic oscillator\index{anharmonic oscillator} 
with the unperturbed
Hamiltonian  $H_0=(p^2+x^2)/2$ 
and the perturbing potential $V=x^4$, 
we may choose $\alpha_k=\sqrt{2^k/k!}$ to get the 
non-vanishing augmented matrix elements from (\ref{HAMPMA1}) and (\ref{HAMPMA2}).
An easy way to calculate the augmented matrix elements starts from the 
ones for $V=x$, where we have
\begin{eqnarray}
\label{HAMPAUG}
W_{k,-1}^x &=& k/2 \, ,\\
W_{k,1}^x &=& 1 \, .
\end{eqnarray}
From here the augmented matrix elements $W_{k,j}^{x^l}$ 
for any polynomial $V=x^l$ can be determined recursively
by applying the rules of matrix multiplication: 
\begin{eqnarray}
\label{HAMPADD}
W_{k,j}^{x^l} = \sum_i W_{k,i}^{x} W_{k+i,j-i}^{x^{l-1}} \,.
\end{eqnarray}
Thus we obtain for $V=x^4$:
\begin{eqnarray}
\label{HAMPWNK}
W_{k,-4}^{x^4} &=& k(k-1)(k-2)(k-3)/16 \, , \\
W_{k,-2}^{x^4} &=& k(2k-1)(k-1)/4 \, ,   \\
W_{k,0}^{x^4}  &=& 3 (2 k^2+2 k+1)/4 \, ,  \\
W_{k,2}^{x^4}  &=& 2k+3 \, ,          \\
W_{k,4}^{x^4}  &=& 1        \, .      
\end{eqnarray}
Being rational numbers they are suitable for recursive calculations even 
up to very high orders. In Tab.~\ref{HAMPTabIV} we have listed
the first ten coefficients $\epsilon_k^{(0)}$ for 
the expansion of the ground-state energy $E_0(g)$.
\pagebreak
\begin{table}[t]
\caption[TabIV]{\label{HAMPTabIV} The first ten coefficients $\epsilon_k^{(0)}$
of the weak-coupling expansion\index{weak-coupling!expansion}\index{expansion!weak-coupling} 
for the ground-state energy\index{ground-state!energy}\index{energy!ground-state} $E_0(g)$ of 
the anharmonic oscillator\index{anharmonic oscillator}\index{oscillator!anharmonic}.}
\begin{center}
\begin{minipage}{5cm}
\begin{tabular}{|c|c|}\hline 
\multicolumn{2}{|c|}{\bf 
The Anharmonic Oscillator\index{anharmonic oscillator}\index{oscillator!anharmonic}} \\ \hline
$k $ & $\epsilon_k^{(0)}$  \\ \hline \hline
$0 $ & $ 1/2 $ \\ \hline
$1 $ & $ 3/4 $ \\ \hline
$2 $ & $ -21/8 $ \\ \hline
$3 $ & $ 333/16 $ \\ \hline
$4 $ & $ -30\,885/128 $ \\ \hline
$5 $ & $ 916\,731/256 $ \\ \hline
$6 $ & $ -65\,518\,401/1\,024 $ \\ \hline
$7 $ & $ 2\,723\,294\,673/2\,048 $ \\ \hline
$8 $ & $ -1\,030\,495\,099\,053/32\,768 $ \\ \hline
$9 $ & $ 54\,626\,982\,511\,455/65\,536 $ \\ \hline
$10 $ & $ -6\,417\,007\,431\,590\,595/262\,144 $ \\ \hline 
\end{tabular}
\end{minipage}
\end{center}
\end{table}
\section*{Acknowledgement}
It is a particular pleasure for B.H. to contribute to this volume in 
honor of Hagen Kleinert, looking back gratefully on four decades of
friendship originating in the physics undergraduate courses of 
Technische Universit\"at Hannover, 
incorporating a common year at Boulder University
as well as the last 32 years as colleagues together at the Freie Universit\"at
Berlin. This is written with great admiration for Hagen's deep dedication
to the miracles of nature and their physical interpretation.
\end{document}